\documentclass[aps,prx,twocolumn,superscriptaddress,longbibliography]{revtex4-2}
\usepackage{color}
\usepackage{dcolumn}
\usepackage{graphicx}
\usepackage{amsmath}
\usepackage{amssymb}
\usepackage{bm}
\usepackage{braket}
\usepackage[hidelinks]{hyperref}

\usepackage{soul}
\usepackage{xcolor}
\begin{document}

\title{Scalable Low-overhead Superconducting Non-local Coupler with Exponentially Enhanced Connectivity}

\author{Haonan Xiong}
\thanks{These authors contributed equally to this work.}
\affiliation{Center for Quantum Information, Institute for Interdisciplinary Information Sciences, Tsinghua University, Beijing 100084, China}
\affiliation{Beijing Academy of Quantum Information Sciences, Beijing 100193, China}
\author{Jiahui Wang}
\thanks{These authors contributed equally to this work.}
\affiliation{Center for Quantum Information, Institute for Interdisciplinary Information Sciences, Tsinghua University, Beijing 100084, China}
\author{Juan Song}
\thanks{These authors contributed equally to this work.}
\affiliation{Beijing Academy of Quantum Information Sciences, Beijing 100193, China}
\affiliation{Institute of Physics, Chinese Academy of Science, Beijing 100190, China}
\affiliation{University of Chinese Academy of Sciences, Beijing 101408, China}
\author{Jize Yang}
\affiliation{Center for Quantum Information, Institute for Interdisciplinary Information Sciences, Tsinghua University, Beijing 100084, China}
\author{Zenghui Bao}
\affiliation{Center for Quantum Information, Institute for Interdisciplinary Information Sciences, Tsinghua University, Beijing 100084, China}
\author{Yan Li}
\affiliation{Center for Quantum Information, Institute for Interdisciplinary Information Sciences, Tsinghua University, Beijing 100084, China}

\author{Zhen-Yu Mi}
\affiliation{Beijing Academy of Quantum Information Sciences, Beijing 100193, China}

\author{Hongyi Zhang}
\affiliation{Center for Quantum Information, Institute for Interdisciplinary Information Sciences, Tsinghua University, Beijing 100084, China}
\affiliation{Hefei National Laboratory, Hefei 230088, China}

\author{Hai-Feng Yu}
\affiliation{Beijing Academy of Quantum Information Sciences, Beijing 100193, China}
\affiliation{Hefei National Laboratory, Hefei 230088, China}

\author{Yipu Song}\email{ypsong@mail.tsinghua.edu.cn}
\affiliation{Center for Quantum Information, Institute for Interdisciplinary Information Sciences, Tsinghua University, Beijing 100084, China}
\affiliation{Hefei National Laboratory, Hefei 230088, China}

\author{Luming Duan}\email{lmduan@tsinghua.edu.cn}
\affiliation{Center for Quantum Information, Institute for Interdisciplinary Information Sciences, Tsinghua University, Beijing 100084, China}
\affiliation{Hefei National Laboratory, Hefei 230088, China}

\date{\today}

\begin{abstract}
Quantum error correction codes with non-local connections such as quantum low-density parity-check (qLDPC) incur lower overhead and outperform surface codes on large-scale devices. These codes are not applicable on current superconducting devices with nearest-neighbor connections. To rectify the deficiency in connectivity of superconducting circuit system, we experimentally demonstrate a convenient on-chip coupler of centimeters long and propose an extra coupler layer to map the qubit array to a binary-tree connecting graph. This mapping layout reduces the average qubit entangling distance from $O(N)$ to $O(\log N)$, demonstrating an exponentially enhanced connectivity with eliminated crosstalk. The entangling gate with the coupler is performed between two fluxonium qubits, reaching a fidelity of 99.37 \% while the system static ZZ rate remains as low as 144 Hz without active cancellation or circuit parameter targeting. With the scalable binary tree structure and high-fidelity non-local entanglement, novel quantum algorithms can be implemented on the superconducting qubit system, positioning it as a strong competitor to other physics systems regarding circuit connectivity.
\end{abstract}

\maketitle

\section{Introduction}
Quantum circuit connectivity is gaining increasing importance for breakthrough from the NISQ era to the fault-tolerant era. With connection beyond nearest-neighbor entanglement, new quantum error correction codes can be implemented, as demonstrated in neutral-atom and trapped-ion systems\cite{bluvstein_logical_2024,reichardt_demonstration_2024}. Recently, a growing interest in quantum low-density parity-check (qLDPC) codes has been motivated by its low overhead and high threshold\cite{delfosse_bounds_2021,cohen_low-overhead_2022,tremblay_constant-overhead_2022,strikis_quantum_2023,bravyi_high-threshold_2024,xu_constant-overhead_2024}. One problem of traditional surface codes is its low encoding rate. The number of physical qubits of one logical qubit scales quadratically with the code distance\cite{fowler_surface_2012}. In contrast, the required physical qubit number can be reduced by more than one order of magnitude for large-scale qLDPC codes\cite{xu_constant-overhead_2024}. Superconducting circuit is a system more favorable for nearest-neighbor coupling and surface codes due to its 2D nature. Non-local coupling technology for superconducting circuits is still in its infancy, not to mention the all-to-all connection. As a result, connectivity remains as one of the major shortcomings of superconducting circuits compared to other platforms. 

Non-local entanglement between superconducting qubits is typically mediated with standing wave modes or traveling photons\cite{sillanpaa_coherent_2007,majer_coupling_2007,dicarlo_demonstration_2009,song_10-qubit_2017,kannan_waveguide_2020,cai_all-microwave_2021,niu_low-loss_2023,zhou_realizing_2023,marinelli_dynamically_2023,storz_loophole-free_2023,qiu_deterministic_2023,mollenhauer_high-efficiency_2024,song_realization_2024,deng_long-range_2025}.
 One motivation of the long-range qubit entanglement lies in its potential for modular quantum network or inter-fridge communication\cite{bravyi_future_2022}. In contrast, designing non-local interactions for connectivity enhancement is yet at an early stage. Previous studies in this direction have been focusing on achieving all-to-all connection with a center bus resonator\cite{song_10-qubit_2017,cai_all-microwave_2021,zhou_realizing_2023,marinelli_dynamically_2023}. However, this scheme has several drawbacks on its scalability towards hundreds of qubits. Increasing the bus size along with the qubit number introduces extra bus modes and causes frequency crowding. The inevitable crosstalk will eventually hinder parallel gates between large number of qubit pairs. Another bottleneck is the insufficient gate fidelity due to decoherence and slow entangling rate\cite{zhou_realizing_2023,marinelli_dynamically_2023}. There is so far no demonstration of bus-based gate that is close to the error correction threshold. Thus, it remains highly challenging to realize high-fidelity all-to-all connection on a large scale superconducting device.

In this work, we solve the connectivity problem by taking a step back. While all-to-all connection generally implies the ability to entangle any pair of the qubits with one two-qubit gate, we propose a scheme that can entangle qubit pairs with $O(\log N)$ gates, where $N$ is the total number of qubits. There are two key ingredients in our scheme: two-qubit non-local couplers and binary tree connecting graph. The non-local couplers make it possible to map a 1D qubit chain to a binary tree such that the average number of gates to entangle the qubits reduces from $O(N)$ to $O(\log N)$. Every qubit is allocated with a binary code for identifying the shortest entangling path. Therefore we name this structure a binary entanglement addressing tree (BEAT). By abandoning the central bus design, parallel gates can be executed on different couplers with eliminated crosstalk. The example of BEAT illustrates the power of non-local couplers to construct a scalable and highly connected circuit. The $O(\log N)$-step all-to-all connection makes superconducting circuits competitive compared to ion and atom systems on large scale devices.

We then experimentally demonstrate a prototype device of the non-local coupler with a simple and convenient design. The coupler is a coplanar waveguide (CPW) resonator capacitively coupled to two fluxonium qubits\cite{manucharyan_fluxonium_2009,nguyen_high-coherence_2019,somoroff_millisecond_2023} which is compatible with the flip-chip technique\cite{foxen_qubit_2018} on large-scale devices. This implies that it can realize not only intra-chip connection but also inter-chip connection to extend the size of quantum processors\cite{bravyi_future_2022}. The static ZZ rate is only 144 Hz with no need for active cancellation or accurate circuit parameter targeting\cite{kandala_demonstration_2021,wei_hamiltonian_2022,xiong_arbitrary_2022,li_realization_2024,zhang_tunable_2024,lin_24_2024}. It leads to a remarkably high on-off ratio\cite{mckay_high-contrast_2015} of 2.9 $\times10^4$. The entangling gate is constructed with the geometric phase of the resonator photon after an off-resonant microwave pulse. In this regard, the gate scheme is similar to the resonator-induced-phase (RIP) gate\cite{cross_optimized_2015,paik_experimental_2016,finck_suppressed_2021,malekakhlagh_optimization_2022,kumph_demonstration_2024,deng_long-range_2025}, yet it is implemented with smaller detuning and lower power. This further reduces the crosstalk between parallel gates on different couplers. The CPW resonator is made of high-quality Tantalum film with a quality factor of 8.0 $\times10^5$ and we achieve a CZ gate fidelity of 99.37 \%. The error rate is already at the error correction threshold and further improvements are within reach. The scalability, low crosstalk, high fidelity and long entangling distance make our gate scheme a perfect candidate for the BEAT structure. Introducing our approach to the superconducting circuit toolbox will exponentially enhance the qubit connectivity. 

\begin{figure*}[t]
    \centering
    \includegraphics[width=1\linewidth]{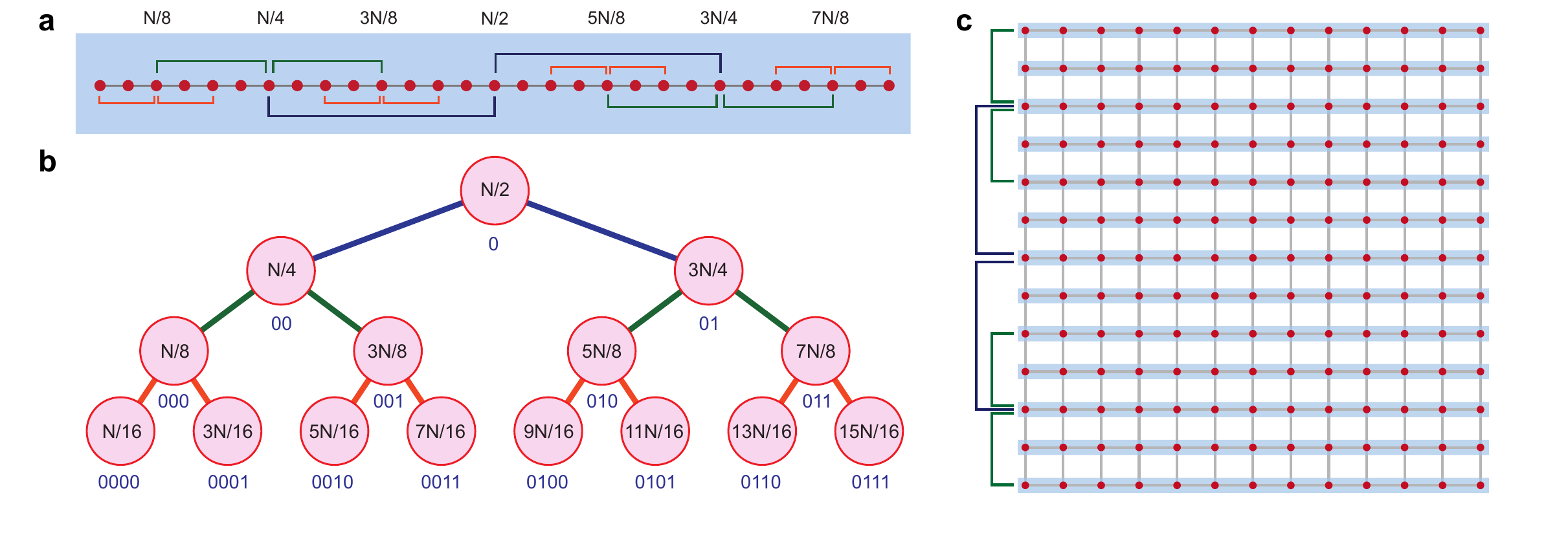}
    \caption{\textbf{Binary entanglement addressing tree (BEAT) for exponentially enhanced connectivity.}
  (a) BEAT design for 1D qubit chain. The brackets represent non-local couplers and the red dots represent qubits. The BEAT assembly can be applied along with the nearest-neighbor coupling. (b) The connection graph of the non-local couplers. The qubits are connected through a binary tree to enhance the connectivity of the graph. The top node is the qubit at the middle of the chain and expand the tree to both directions. The average entangling distance between qubits reduces from $O(N)$ to $O(\log N)$ compared to the 1D chain. The binary address under the qubit is related to the binary representation of the qubit position. (c) BEAT design for 2D grid. Each row of qubits is equipped with 1D BEAT. An extra BEAT assembly improves the connectivity among different rows such that the average entangling distance of the grid also reduces to $O(\log N)$.}
    \label{fig:beat}
\end{figure*}
\section{Binary entanglement addressing tree}
Here we introduce the BEAT layout as an example to demonstrate that non-local couplers can improve the circuit connectivity exponentially. We start from a 1D qubit chain as shown in Fig.~\ref{fig:beat}a. The qubit chain consists of $N-1$ qubits labeled from $1$ to $N-1$. For convenience, we assume $N=2^L$ and $L$ is an integer. The BEAT assembly applies an extra connection layer to the qubit chain, where the qubits are connected through a binary tree demonstrated in Fig.~\ref{fig:beat}b. The root node is qubit $N/2$ located at the center of the chain. Its child nodes are the center qubits of the left part and the right part of the chain. In general, the nodes at level $l$ are located at the positions of $iN/2^{(l+1)}$ and $i=1, 2, ..., l$, excluding the qubits belonging to the upper nodes. The maximum level is $l_\mathrm{max}=L-1$. The connection between the nodes is realized with a non-local coupler.

The longest entangling distance (number of two-qubit gates to entangle two qubits) of a 1D chain with nearest-neighbor coupling is $d=N$. In comparison, the most distant qubit pair in a BEAT layout comes from the two ``leaf'' qubits at the bottom level with the root node as the lowest common ancestor. The entangling distance is only $d=2l_\mathrm{max}=2\log_2 N - 2$. Overall, the average entangling distance between two qubits is $O(N)$ for nearest-neighbor connection layout and $O(\log N)$ for a BEAT layout. This result demonstrates that the non-local couplers have the ability to enhance the circuit connectivity exponentially. All the qubits are ``all-to-all'' connected within $O(\log N)$ steps.

The entangling path between two qubits in this binary tree goes through their lowest common ancestor, which can be identified by assigning an address code to each qubit in the tree as shown in Fig.~\ref{fig:beat}. Every node in BEAT layout can be allocated with a unique binary code. This address code is determined by the binary representation of the qubit position. For instance, the address of position $9N/16$ can be calculated by converting 9/16 to $0.1001_2$ where the subscript indicates that it is a binary fraction. The digits excluding the last one form the address code 0100. The address code of the lowest common ancestor between two qubits is the sequence of matching digits from the start of both qubit address codes. For instance, the matching digits between 0100 and 0110 is 01, indicating that the lowest common ancestor between $9N/16$ and $13N/16$ is $3N/4$.

This ``logarithmic'' all-to-all connectivity can be extended to a 2D grid by applying the BEAT assembly to the rows, as shown in Fig.~\ref{fig:beat}c. The qubits at different rows can be first entangled to their leftmost qubits and then entangled to each other through the row BEAT couplers. The number of steps required in this process is still $O(\log N)$. This example demonstrated that the basic unit in a BEAT layout can not only be a single qubit but also be a group of qubits. The BEAT assembly can be used to scale up quantum modules\cite{niu_low-loss_2023,qiu_deterministic_2023,mollenhauer_high-efficiency_2024,song_realization_2024,deng_long-range_2025} that enables long distance qubit entanglement even beyond meter level.

The BEAT layout demonstrates the power of non-local couplers that brings remarkable improvement on the superconducting circuit connectivity. Compared with a direct all-to-all qubit connection network, our approach is more feasible and scalable. We avoid the frequency crowding problem by distributing the coupler load among all the qubits and relaxing the one-step all-to-all requirement. The total coupler number scales as $O(N)$ and in the worst case, a qubit only couples to three couplers at most. There is no overlap between the couplers, indicating that this can be readily realized on a 2D architecture. The scheme is favorable for parallel gates at different couplers as they are independent modes and the crosstalk can be suppressed with airbridge techniques\cite{chen_fabrication_2014,bu_tantalum_2024}.

\section{Prototype device demonstration\label{sec:proto}}
\subsection{System circuit}
\begin{figure}
    \centering
    \includegraphics[width=86mm]{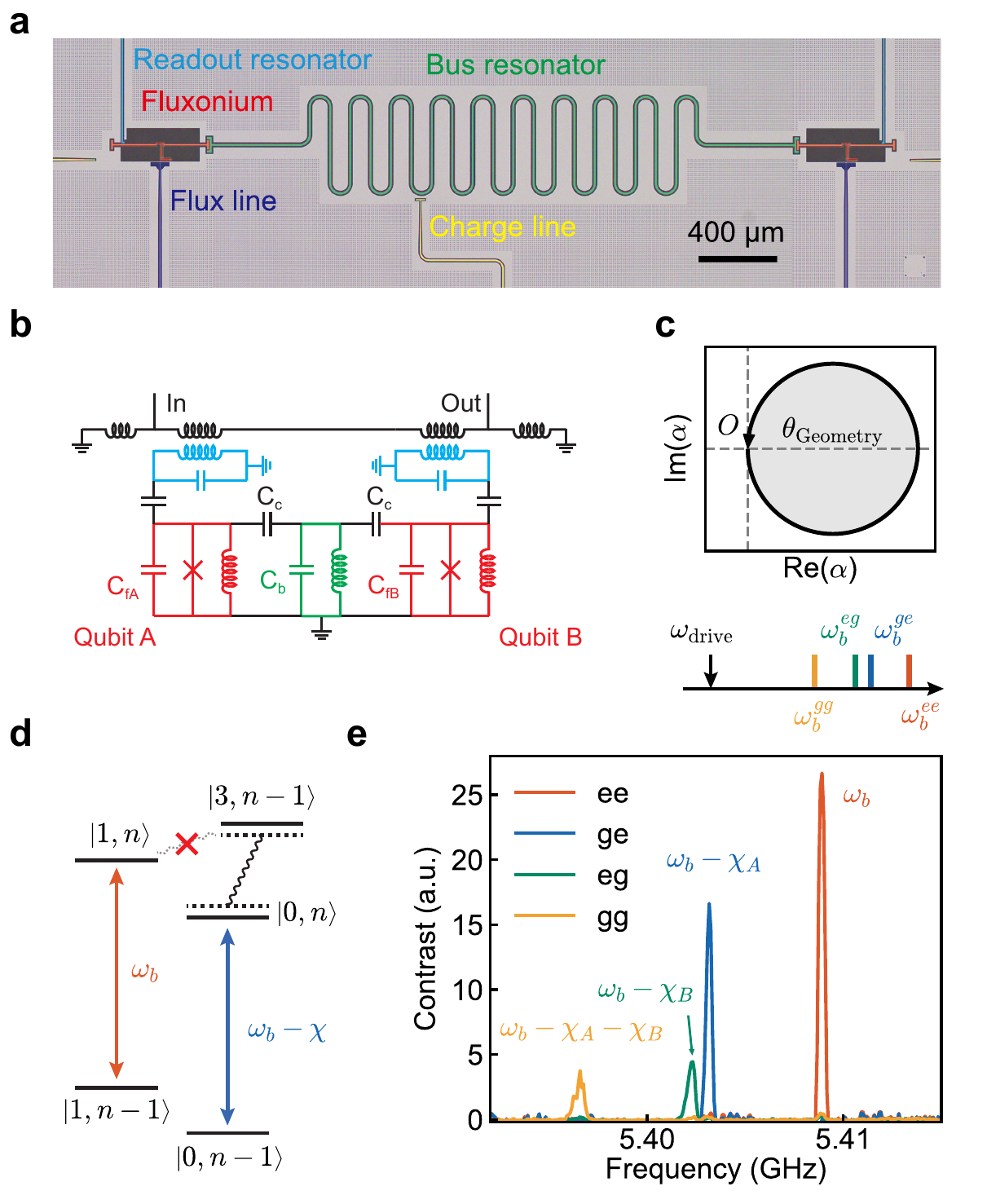}
    \caption{\textbf{Device and gate principle.} (a) False color of the demonstration device. Two fluxoniums (red) are coupled by a bus resonator (green) of 11.4 mm. (b) Circuit schematic. The dedicated readout resonators (cyan) of the fluxoniums are inductively coupled to a shared Purcell filter. (c) Gate principle. Top panel: the evolution of the bus photon state gives rise to a geometric phase defined by the area of the closed path in the resonator phase space. Bottom panel: a detuned drive leads to distinct geometric phases on the four computational states, contributing to a conditional phase. (d) Bus resonator dispersive shift $\chi$ at half flux quanta. $\chi$ mainly comes from the energy repulsion between $\ket{3,n-1}$ and $\ket{0,n}$. There is no interaction between $\ket{3,n-1}$ and $\ket{1,n}$ due to vanishing matrix element at half flux quanta. (e) Measured quartet of the bus in a two-tone experiment when the qubits are prepared at different computational states. The dispersive shifts of the two qubits are $\chi_A=5.4~\mathrm{MHz}$ and $\chi_B=6.9~\mathrm{MHz}$.}
    \label{fig:device}
\end{figure}
Realizing scalable high-fidelity non-local couplers is the cornerstone of the BEAT layout. Here we introduce the design of a non-local coupler that can entangle qubits at centimeter level distance with a gate scheme that fully satisfies the requirements. Our device uses a $\lambda/2$ CPW resonator of 11.4 mm long as the coupler between two fluxonium qubits. This bus resonator capacitively couples to the qubits and a dedicated charge line near the center (Fig.~\ref{fig:device}a). The two fluxoniums both operate at half flux quanta and are provided with their own dedicated flux line and charge line. Each fluxonium is capacitively coupled to a $\lambda/4$ CPW resonator for readout. The readout resonators are inductively coupled to a Purcell filter which is a $\lambda/2$ CPW resonator with both ends shorted to the ground (Fig.~\ref{fig:device}b). The system Hamiltonian can be written as
\begin{multline}
    \hat{H} = \hat{H}_\mathrm{fA}+\hat{H}_\mathrm{fB}+\hat{H}_\mathrm{bus}
    +J_\mathrm{c} \hat{n}_\mathrm{fA} \hat{n}_\mathrm{b} + J_\mathrm{c} \hat{n}_\mathrm{fB} \hat{n}_\mathrm{b}\\+J_\mathrm{AB} \hat{n}_\mathrm{fA} \hat{n}_\mathrm{fB},
    \label{eq:Hamiltonian}
\end{multline}
where $\hat{H}_\mathrm{fA(B)}$ is the fluxonium Hamiltonian, $\hat{H}_\mathrm{bus}$ is the bus Hamiltonian and $\hat{n}_\mathrm{i}$ is the corresponding charge operator. As labeled in Fig.~\ref{fig:device}b, $C_\mathrm{b}, C_\mathrm{fA(B)}$ are the capacitance of the bus and the fluxoniums and $C_\mathrm{fA} = C_\mathrm{fB} = C_\mathrm{f}$. $C_\mathrm{c}$ is the coupling capacitance. Because $C_\mathrm{b}, C_\mathrm{f}>>C_\mathrm{c}$, we obtain $J_\mathrm{c}=4e^2C_\mathrm{c}/(C_\mathrm{b}C_\mathrm{f}), J_\mathrm{AB}=4e^2C_\mathrm{c}^2/(C_\mathrm{b}C_\mathrm{f}^2)$ by keeping the leading order. 

Even though there is no direct capacitance between the two fluxoniums in the circuit model, the capacitance network gives rise to an effective capacitive coupling $J_\mathrm{AB}$. For our device, $C_\mathrm{c}=2.2~\mathrm{fF}, C_\mathrm{c}/C_\mathrm{f}=0.11, C_\mathrm{c}/C_\mathrm{b}=3.7\times10^{-3}$. These designed parameters lead to an order of magnitude difference between $J_\mathrm{AB}$ and $J_\mathrm{c}$. A small $J_\mathrm{AB}$ ensures suppressed qubit crosstalk as it can contribute to mode hybridization and static ZZ interaction. Both $J_\mathrm{AB}$ and $J_\mathrm{c}$ are the source of static ZZ in our system. Because $J_\mathrm{AB}$ is a direct coupling between the two qubits, it has a comparable effect on static ZZ (with an opposite sign) despite the fact that $J_\mathrm{c}$ is much larger. This leads to a natural ZZ suppression in the coupled system\cite{kandala_demonstration_2021,ding_high-fidelity_2023}. With $J_\mathrm{c}/h=30~\mathrm{MHz}, J_\mathrm{AB}/h=3.4~\mathrm{MHz}$, the static ZZ can be calculated by diagonalizing the system Hamiltonian $\zeta_\mathrm{ZZ}/h=1.7~\mathrm{Hz}$. We measure the static ZZ by performing Ramsey experiments on qubit A as shown in Appendix \ref{sec:staticZZ}. The measured static ZZ rate is 144 Hz. The discrepancy is possibly caused by stray capacitance on the chip and the systematic measurement error on the ZZ rate that is much slower than the system decoherence. Such a small static ZZ rate is negligible in gate operation, demonstrating the natural suppression on the ZZ crosstalk, which is beneficial in large-scale devices.
\subsection{Gate principle \label{sec:gate_principle}}

By driving the bus resonator, we can realize a CZ gate through the geometric phase from the photon state evolution. As shown in Fig.~\ref{fig:device}c, after the photon state goes through a closed path in the phase space, it acquires a geometric phase that is proportional to the area enclosed by the loop. The bus resonator is coupled to the two fluxoniums with dispersive shifts $\chi_A$ and $\chi_B$. Therefore the detuning from the bus transition is relying on the qubit state, leading to distinct evolution trajectories. This eventually yields a conditional phase that can be used to construct a CZ gate.

The dispersive shift between the bus and fluxonium at half flux quanta mainly comes from the repulsion between the resonator and the $\ket{3}$ states of the fluxonium. In Fig.~\ref{fig:device}d, we use $\ket{k,n}$ to label the system levels, where $k$ is the fluxonium level index and $n$ is the resonator level index. State $\ket{0,n}$ and $\ket{3,n-1}$ repel each other. Thus, the resonator transition $\ket{0,n}-\ket{0,n-1}$ is shifted by $\chi$. On the other hand, although state $\ket{1,n}$ is closer to $\ket{3,n-1}$ in frequency, the coupling term does not hybridize these two states due to parity mismatch at fluxonium sweet spot. As a result, the resonator transition is not pushed by the fluxonium $\ket{3}$ state when it is at $\ket{1}$ state. Relevant transition frequencies of the system are listed in Table~\ref{tab:dispersive_transitions}. In our experiment, we observe that the readout resonators response to the photons in the bus resonator. This enable us to directly measure the bus transitions with conventional two-tone experiments, as shown in Fig.~\ref{fig:device}e. We prepare the two qubits at the four computational states (i.e., gg, ge, eg and ee) and identify the bus transition in each case. The dispersive shifts of the two qubits on the bus are $\chi_A=5.4~\mathrm{MHz}$ and $\chi_B=6.9~\mathrm{MHz}$.

\begin{center}
\begin{table}
\renewcommand{\arraystretch}{1.5}
\setlength{\tabcolsep}{5pt}
\begin{tabular}{  c | c  c  c  c  c  } \hline
    transition & $f_{03}^A$ & $f_{03}^B$ &  $f_{13}^A$ & $f_{13}^B$ & $f_\mathrm{b}$                      \\ \hline
    frequency (GHz)  & 5.748 &5.724&5.400&5.457  & 5.409 \\ \hline
\end{tabular}
\caption{\label{tab:dispersive_transitions} Transitions near the bus frequency. We list the 0-3 and 1-3 transition frequencies for both qubits at sweet spot and the bus frequency $f_b$. The 1-3 transition dose not contribute to the dispersive shifts regardless of the smaller detunings.}
\end{table}
\end{center}

The photon-induced geometric phase gate in our entanglement protocol is similar to traditional RIP gate. They both eventually result in a CZ gate by off-resonantly driving the bus quartets from the two fluxoniums' dispersive shifts. RIP gate typically requires large detuning and strong power to satisfy adiabatic condition and ensure gate speed. This brings the concern of driving unwanted transitions and heating issue, especially on a large-scale device. On the contrary, we adopt non-adiabatic driving scheme, meaning that we can entangle the two qubits with a smaller detuning and weaker power, which ensures the scalability of our scheme.

\subsection{Photon number calibration \label{sec:photon_cal}}

\begin{figure}
    \centering
    \includegraphics[width=86mm]{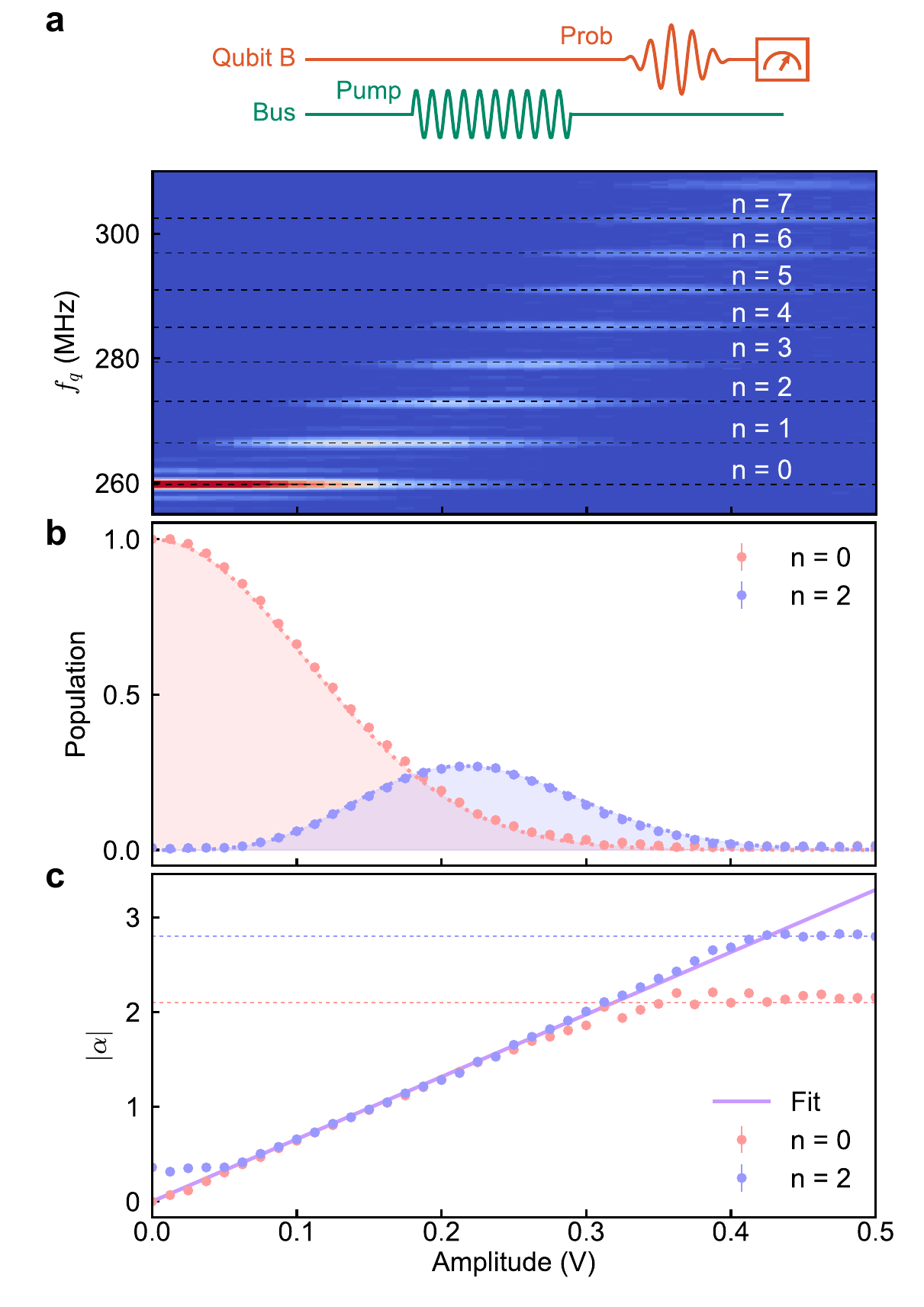}
    \caption{\textbf{Bus photon state calibration using qubit B.} (a) Photon-number-splitting of qubit B transition. A pump pulse is sent to the bus to displace the photon state followed by a prob pulse on qubit B to detect its frequency. The Fock state population of the bus photon splits the qubit B transition as the pump pulse amplitude increases. (b) Fock state population at $n=0$ and $n=2$. The Fock state population can be computed by measuring the qubit transition peak height. The dashed lines are the expected population from the coherent state assumption. (c) Coherent state amplitude $|\alpha|$. The measured Fock state population can be converted to $|\alpha|$ when it is significantly larger than zero. Converted $|\alpha|$ using $n=0$ and $n=2$ Fock state is in agreement and both change linearly with pump pulse amplitude. The purple solid line is a fitting line using the $|\alpha|$ values in the linear regime. The $|\alpha|$ conversion breaks down when the Fock state population drops to the level of 0.01, e.g., when $|\alpha|>2.1$ ($n=0$) and $|\alpha|>2.8$ ($n=2$). This sets a measurement limit on the coherent state amplitude. The error bars are the standard errors.}
    \label{fig:bus_photon}
\end{figure}

To study the dynamics of the bus photon, we calibrate the photon state by measuring the photon-number-splitting peak height of qubit B. We inject photon to the bus resonator with a microwave pump pulse. In the case of weak coupling and low photon number, we can treat the bus resonator as a linear system. For a linear drive,
\begin{equation}
    \hat{V}_\mathrm{drive}(t)=\hbar\Omega(t)(\hat{a}^\dag+\hat{a}).
    \label{eq:bus_drive}
\end{equation}
The photon state evolves in the form of a coherent state and the average photon number is proportional to the amplitude of pump pulse. This process is followed by a measurement of qubit B transition frequency. As depicted in Fig.~\ref{fig:bus_photon}a, we observe qubit B's response at $f_{01}^B+n\chi~(n=0,1,2...)$ with increasing pulse amplitude, corresponding to the bus photon at Fock state $\ket{n}$. Because in our experiment the readout resonator also responses to the bus photons, the measurement background drifts with the pump pulse amplitude. We subtract this drifting background to better visualize the qubit photon-number-splitting peaks.

The height of photon-number-splitting peaks is proportional to the photon Fock state population $P$. We measure the Fock $\ket{0}$ state peak height by applying a $\pi$ pulse at the corresponding frequency and comparing the readout signal to the measurement without a $\pi$ pulse. This gives voltage difference $V_{0\pi}(P)=|V_0(P)-V_\pi(P)|=kP$. We assume the Fock $\ket{0}$ state population is 1 when no photon is injected to the bus, i.e., the pump amplitude is 0. In this case, the voltage difference $V_{0\pi}(1)=k$. Hence the Fock $\ket{0}$ state population $P=V_{0\pi}(P)/V_{0\pi}(1)$ and the corresponding coherent state amplitude $|\alpha|=\sqrt{-\log(P)}$. Similarly, we can also calculate the coherent state amplitude using the Fock $\ket{2}$ state peak height. The Fock $\ket{2}$ state population can be calibrated with the help of Fock $\ket{0}$ state. We first pick the data points at amplitude of 0.23 V and calculate $|\alpha|$ with Fock $\ket{0}$ state population. Then we calculate Fock $\ket{2}$ state population with $|\alpha|$, which leads to the coefficient $k$ for Fock $\ket{2}$ state peak height thereafter. As demonstrated in Fig.~\ref{fig:bus_photon}c, the coherent state amplitude $|\alpha|$ calculated using Fock $\ket{0}$ state and Fock $\ket{2}$ state is in agreement at other pump amplitudes. $|\alpha|$ increases linearly with pump amplitude, which is consistent with the linear system assumption. The reliability of this calculation is dependent upon the Fock state population. When the Fock state population drops to the level of 0.01, systematic errors from thermal excitation, incoherent process and nonlinearity dominate and the conversion to $|\alpha|$ is not accurate. For this reason, $|\alpha|$ saturates when $|\alpha|>2.1$ calculated from Fock $\ket{0}$ state population and when $|\alpha|>2.8$ or $|\alpha|<0.4$ calculated from Fock $\ket{2}$ state population. We choose a regime where $|\alpha|$ follows a linear trend and fit for the ratio between $|\alpha|$ and pump amplitude. This ratio is used to calculate the Fock state population distribution (dotted line in Fig.~\ref{fig:bus_photon}b), showing good agreement across the whole amplitude range.

\subsection{Gate performance\label{sec:gate}}

In order to obtain a high fidelity gate, the most critical point is to guide the bus photon back to the vacuum state for all the four computational states at the end of the gate, which requires designing the drive pulse carefully\cite{cai_all-microwave_2021}. For a harmonic oscillator and a drive of time $T$ with a complex amplitude $\Tilde{\Omega}(t)$, the coherent state response is\cite{cross_optimized_2015}
\begin{equation}
    \alpha(T)=-\frac{i}{2}e^{-i\Delta T}\int_0^Te^{-i\Delta t}\Tilde{\Omega}(t) \mathrm{d}t,
    \label{eq:alpha_evolv}
\end{equation}
where $\Delta$ is the detuning of the drive. Eq.~\ref{eq:alpha_evolv} indicates that the final coherent state amplitude $|\alpha(T)|$ is the Fourier amplitude of the complex drive at the detuning frequency $|F_{\Tilde{\Omega}}(\Delta)|$. In our system, the detuning of the drive is dependent on the qubit state due to dispersive shift, i.e., $\Delta=\Delta_{ij}~(i,j=0,1)$. Hence the drive should satisfy the condition $|F_{\Tilde{\Omega}}(\Delta_{ij})|=0$ for $i,j=0,1$. We perform the required pulse shaping with an iterative spectrum engineering (ISE) method presented in Appendix \ref{sec:pulse_opt} where we use four free parameters to control the four zero points of the pulse Fourier amplitude. In practice, the bus resonator gains anharmonicity from the coupling to the fluxoniums at high photon number. Therefore we observe that Eq.~\ref{eq:alpha_evolv} deviates from the actual situation, especially at high power. In experiment, we optimize the pulse parameters according to the measurement and use Nelder-Mead algorithm to minimize the residual photons.

After optimizing the shape of a 120-ns pulse and drive frequency, we can suppress the average residual photon number down to 1e-2 level, as demonstrated in Fig.~\ref{fig:CZ_gate}a. This is consistent with the optimal residual photon in our simulation in Appendix~\ref{sec:error}. We measure the photon number evolution during the pulse. It shows good agreement with the simulation in the case for the four computational states (Fig.~\ref{fig:CZ_gate}b). Because the drive frequency is on the low frequency side that is closer to the gg state, the photon number is the highest when the qubits are at gg state. In Fig.~\ref{fig:CZ_gate}c, we measure the optimized CZ gate fidelity with randomized benchmarking (RB). The fidelity reaches 99.48$\pm$0.06 \%. To further confirm the gate fidelity and analyze the error source, we interleave multiple CZ gates in RB sequence in Fig.~\ref{fig:CZ_gate}d. We fit the error rate $\epsilon(m)=1-F(m)$ versus gate number $m$ using a quadratic model $\epsilon(m)=\epsilon_1 m+\epsilon_2 m^2$. The linear term is dominated by decoherence and the quadratic term is dominated by coherent errors\cite{sheldon_characterizing_2016}, which possibly originates from residual photons. We obtain $\epsilon_1=0.52~\%,\epsilon_2=0.11~\%$, indicating the error rate of a single CZ gate is $\epsilon(1)=0.63~\%$. We estimate the error budget in Appendix~\ref{sec:error} with a total error of 0.54~\% and conclude that the major error arises from the dephasing of the qubits and the bus photon. Because the photon number is relatively low during the gate pulse, the measurement-induced state transition\cite{boissonneault_nonlinear_2008,sank_measurement-induced_2016,khezri_measurement-induced_2022,nesterov_measurement-induced_2024,dumas_unified_2024} is not prominent in the gate dynamics, although it can be commonly observed in readout process (Appendix~\ref{sec:readout}). We prove this by simulating the population evolution during the gate. This suggests that the gate error can be corrected with standard procedures of error correction codes.

\begin{figure}
    \centering
    \includegraphics[width=86mm]{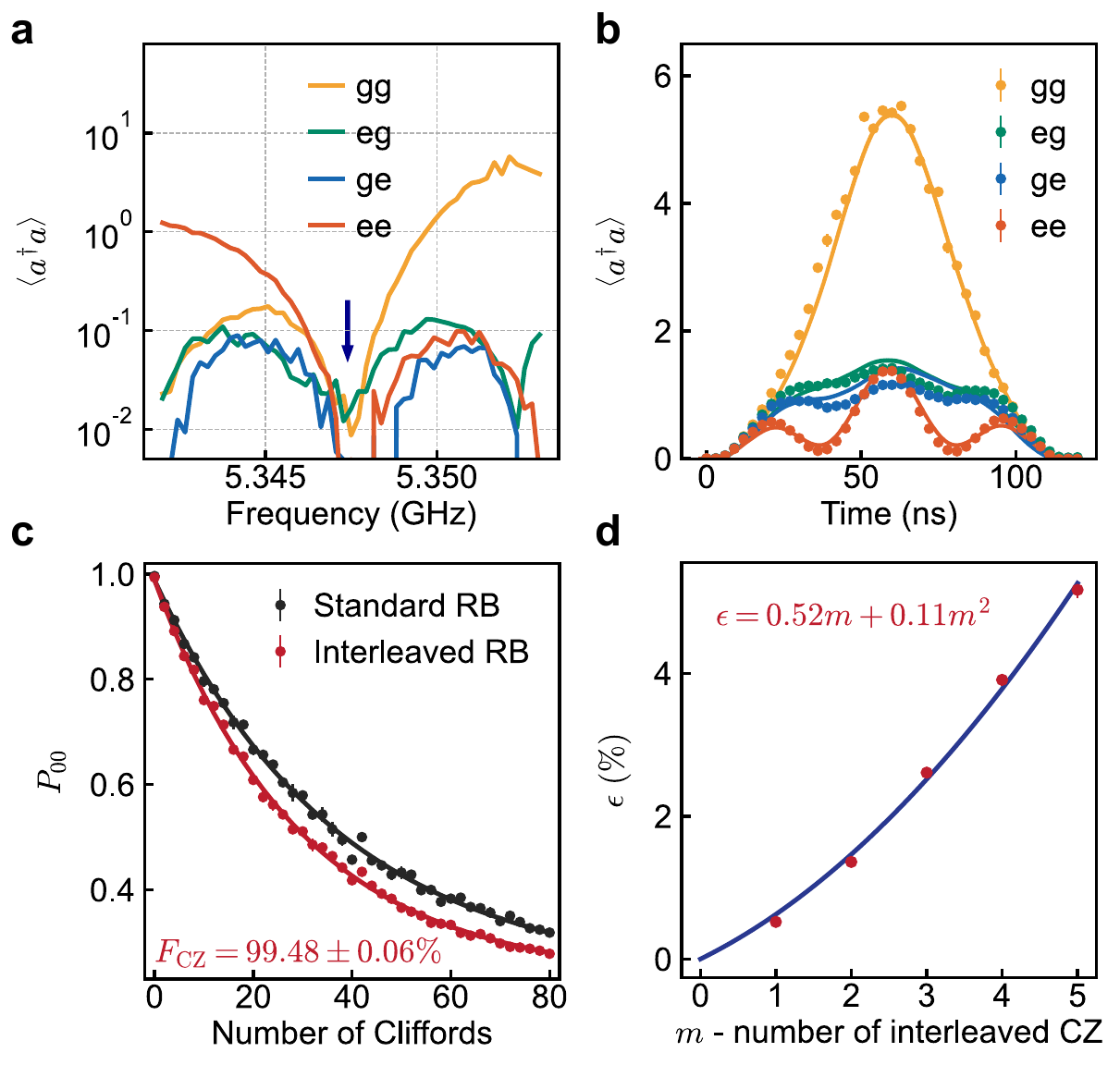}
    \caption{\textbf{CZ gate optimization and benchmarking} (a) Residual photon number $\braket{a^\dagger a}$ at different drive frequencies measured with the protocol in Sec.~\ref{sec:photon_cal}. The blue arrow indicates the optimal drive frequency. (b) Photon number during the CZ gate pulse for different computational states. The measurement (dots) and the simulation (lines) are in good agreement. (c) CZ gate Randomized benchmarking. The fidelity of a single CZ gate is 99.48$\pm$0.06 \%. (d) Error rate of $\mathrm{CZ}^m$ versus $m$. The errors are fitted with a quadratic model. The quadratic dependence on $m$ indicates a coherent part in the gate error, which can be attributed to the residual photons at the end of the gate. The error bars in this figure are the standard errors.}
    \label{fig:CZ_gate}
\end{figure}

\section{Discussion and outlook}
The fluxonium capacitance eventually limits the total coupling capacitance that can be loaded onto the qubit. From this perspective, the BEAT layout takes advantage of distributing the capacitance load among qubits with no more than three non-local couplers on one qubit. Its low overhead makes it a practical and scalable tool for ``logarithmic'' all-to-all connectivity. Fixed capacitive coupling eliminates the need for flux lines and makes our non-local coupler compatible with flip-chip technique. When the control lines are on the same chip with the bus, they can step over the resonators at the nodes of the standing wave modes using airbridges to suppress the crosstalk. 

One advantage of the BEAT layout over the center bus design is the ability to execute parallel gates. However, malfunctioning qubits can be a potential concern in large scale superconducting circuits. Two-level systems or broken wiring can prevent a node qubit from entangling two children qubits. One solution is to embed two or more qubits at each parent node (especially the root node) to ensure the robustness of the connecting graph. While the binary tree structure is suitable for a 1D chain, one can also adopt $d$-ary tree in general. In practice, it is more efficient to design the connecting graph according to the requirement of a specific algorithm such as qLDPC.

 The exponential enhancement on connectivity with the BEAT layout makes superconducting circuits competitive compared to atom or ion systems. One popular technique in these systems to achieve all-to-all connection is to transport the atoms or ions with the electric field\cite{pino_demonstration_2021} or optical tweezers\cite{bluvstein_quantum_2022}. Nevertheless, the shuttling speeds in these systems are limited to maintain high operation fidelities. Therefore the decoherence error from the idling time in a long-distance transportation will eventually be the bottleneck of the shuttling process with a scaling of $O(N)$. In contrast, the $\log N$ dependence of the BEAT layout has the potential to showcase its strengths for a large number of qubits. To unleash the full potential of BEAT layout, the non-local coupler length should be further extended and the entangling gate fidelity should be approximately one order of magnitude lower than the error-correction threshold.

Since meter-level remote entanglement has been demonstrated with tunable inductive couplers\cite{qiu_deterministic_2023,niu_low-loss_2023}, here we discuss methods to extend the length of the bus resonator with fixed capacitive coupling. Apart from the CPW fundamental mode used in Sec.~\ref{sec:proto}, one can also use higher harmonic modes such that the bus length can be doubled or tripled to maintain the same mode frequency. Another possible solution is to split the CPW to three coupled sections of the same length. The hybridized mode can be used as the coupler mode\cite{mckay_high-contrast_2015}. In addition to the fluxonium $\ket{3}$ state, the $\ket{2}$ state can be coupled to the bus as well. This can lower the bus frequency down to 3 GHz level, allowing for a longer CPW resonator. 

The advantage of small drive detuning and weak power differentiates our gate scheme from traditional RIP gate. This is beneficial regarding to suppressed crosstalk as the bus charge line can couple to other fluxonium transitions or couplers due to stray coupling and mode hybridization. The gate fidelity can be improved from the following perspectives. Firstly, larger coupling helps to reduce gate time and the photon number during the gate, which suppresses the incoherent error. There is a competing effect from the dephasing induced by the $\ket{3}$ states, which needs to be taken into account in optimizing coupling strength. In RIP gate scheme, a large detuning with strong power can suppress the measurement-induced dephasing error of the gate\cite{gambetta_qubit-photon_2006,cross_optimized_2015}. However, this effect is not observed in our system, probably also due to the photon dephasing caused by fluxonium $\ket{3}$ states. The $\ket{3}$ state coherence is likely limited by the dielectric loss from the Josephson junction chain. Hence better inductance fabrication methods\cite{grunhaupt_granular_2019,rieger_granular_2022,gupta_low_2024} should improve this situation. On the other hand, the idea of introducing anharmonicity to the bus resonator\cite{rosenfeld_designing_2024} can restrict the photon dynamics within the lowest two levels. This simplified system is beneficial for residual photon suppression.

Our experiment demonstrates fluxonium readout performance close to that of transmon, with binary readout fidelity\cite{riste_initialization_2012} of 98.16~\% and QND-ness\cite{braginsky_quantum_1996,touzard_gated_2019,hazra_benchmarking_2024} of 97.28~\%(Appendix~\ref{sec:readout}). To exclude the assignment errors, especially from the higher state leakage, we extract the population from circular thresholds. In this case, we improve the readout fidelity to 98.91~\% and QND-ness to 98.40~\%. Apart from the high coherence and large anharmonicity, weak static ZZ is another advantage of fluxonium qubits. The static ZZ of 144 Hz corresponds to an on-off ratio of $2.9\times10^4$, which is achieved without tuning flux or precisely targeting circuit parameters. Therefore, we prove that fluxonium can be a good candidate as the building block of large scale superconducting circuits.

\section{Conclusions}
We have proposed a straightforward design of on-chip non-local couplers with capacitively coupled CPW resonators. We have demonstrated that the non-local coupler is able to enhance the connectivity exponentially through the BEAT layout. It is a feasible and scalable approach to reduce the average entangling distance from $O(N)$ to $O(\log N)$ with suppressed crosstalk. We have shown that the non-local coupler can entangle two qubits at a distance of 11.4 mm with a CZ gate fidelity of 99.37 \%. The static ZZ rate is naturally suppressed to 144 Hz without active cancellation or circuit parameter targeting, corresponding to a remarkably high on-off ratio of 2.9 $\times10^4$. Our proposal has made a breakthrough in solving the connectivity problem of superconducting circuits and demonstrated ``logarithmic'' all-to-all connectivity. Our non-local coupler design opens the door for executing new types of error correction codes like qLDPC on superconducting platform.

\begin{acknowledgements}
This work is supported by Innovation Program for Quantum Science and Technology (Grant No.2021ZD0301704, No.
2021ZD0301802), the National Natural Science Foundation of China (Grant No. 12374472, Grant Nos. 92365206), China Postdoctoral Science Foundation (Grant No.2023M742005) and Shuimu Tsinghua Scholar Program. 
\end{acknowledgements}

\appendix

\section{Experimental Setup \label{sec:setup}}

\begin{figure*}[t]
    \centering
    \includegraphics[width=0.97\linewidth]{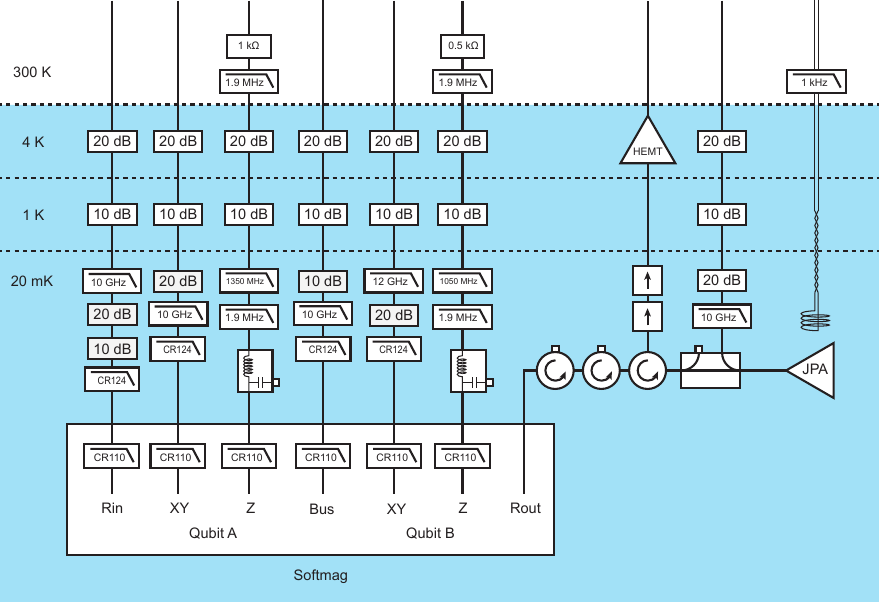}
    \caption{\textbf{Cryogenic wiring.} The filters and attenuators are shown as the labels in the box.}
    \label{fig:wiring}
\end{figure*}

The cryogenic wiring diagram is shown in Fig.~\ref{fig:wiring}. The device is measured at the base plate in an Oxford Triton 400 dilution refrigerator of 10 mK. All the input lines are heavily filtered to suppress noise from the thermal environment. The output signal is amplified with a home-made JPA biased by a coil before going to the HEMT. The sample box is magnetically shielded by a can made of Softmag 4K alloy. We also add another double-layer mumetal magnetic shielding can outside the fridge for extra protection. The readout and control signals are generated with Sinolink SLFS0218H, Agilent N5183A, Agilent N5183B and R\&S SMBV100B and are modulated with room temperature IQ mixers and RF switches. The initialization tones are amplified with room temperature amplifiers to pump the transitions efficiently. All the room temperature electronics are powered by nicslab XDAC-40MUB-R4G8. The control pulses are generated with Tektronics AWG5014C. The fluxes of the fluxonium and JPA are controlled with YOKOGAWA GS210.

\section{Fabrication \label{sec:fab}}

The fabrication process is as follows. The sample is fabricated on a 2-inch c-plane sapphire substrate with both-sides polished. The substrate is annealed up to 1100 °C prior to the deposition of 200 nm alpha-tantalum using DC sputtering. Qubit pads, readout resonators, and transmission lines are patterned using direct write lithography (by Heidelberg DWL $66^+$) with S1813 resist and developed with MF 319 developer. CF$_4$ gas is employed for dry etching. The residual chemicals caused by dry etching is removed by oxygen plasma ashing for 5 min. After dicing the wafer covered with S1813 photoresist, we soak the chip in NMP overnight and then heat it to 105 °C for three hours, followed by a TAMI cleaning process\cite{place_new_2021}. Next, the organic residuals and the oxide layer are removed in a 2:1 H$_2$SO$_4$:H$_2$O$_2$ piranha solution (60 °C for 20 mins) and 10:1 buffered oxide etch (BOE) for 20 mins. The chip is coated with MMA EL 11 resist, PMMA A3 resist and a 13-nm Al anti-charging layer before e-beam lithography. After writing with a RAITH150 e-beam writer, the anti-charging layer is removed with MF319. The device is developed in MIBK:IPA 3:1 and rinsed in IPA. The fluxonium is fabricated with a double-angle deposition process after Argon ion beam etching on the substrate. The liftoff is done in remover PG after 85 °C heating for an hour. The device is then wire-bonded in a Al sample box and loaded into the fridge.

\section{Static ZZ interaction\label{sec:staticZZ}}
\begin{figure}
    \centering
    \includegraphics[width=86mm]{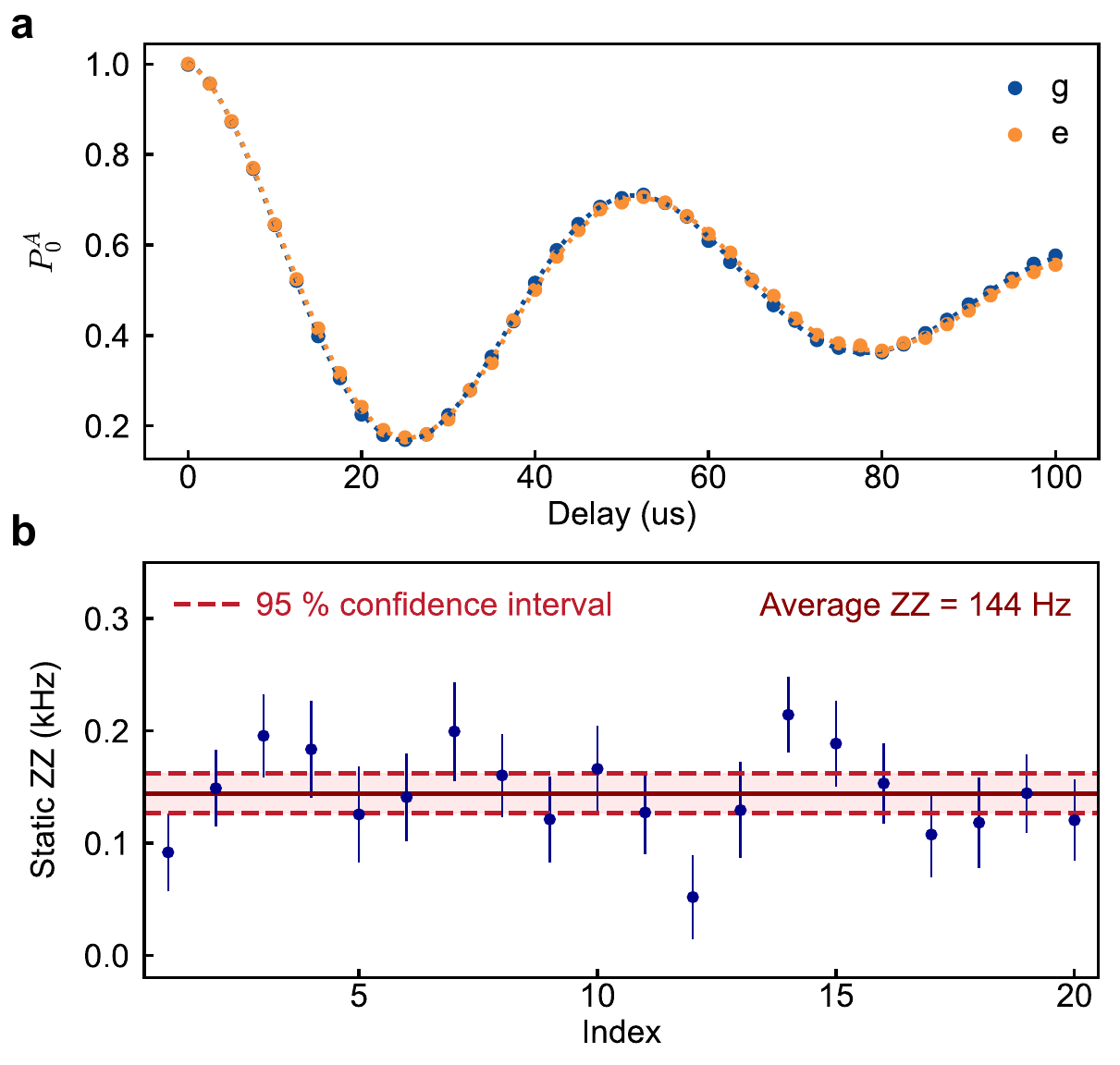}
    \caption{\textbf{Measuring static ZZ interaction from Ramsey experiments} (a) Ramsey experiment on qubit A when qubit B is at different states. The two curves are almost identical, indicating an insignificant ZZ strength. (b) Repetitive static ZZ measurements by fitting the two Ramsey curve in (a). Due to the limited coherence, the static ZZ rate of a single measurement is not accurate. Therefore we repeat the measurement for 20 times and obtain an average ZZ of 144 Hz with a 95 \% confidence interval of [127, 162] Hz. The error bars in the figure are the standard errors.}
    \label{fig:ZZ}
\end{figure}

One of the advantages of our resonator microwave coupler is the high on-off ratio. A fast entangling gate can be implemented when the coupler is turned on with a driving pulse. In contrast, the two qubits barely entangle when no microwave is applied to the coupler. In this case, the major entangling effect is the static ZZ interaction. As shown in Fig.~\ref{fig:ZZ}a, we measure the Ramsey oscillation of qubit A when qubit B is prepared at g state and e state respectively. Despite the subtle difference between the two Ramsey curves, we extract the static ZZ rate by fitting for the oscillation frequency and repetitive measurements. The average static ZZ rate from 20 measurements is 144 Hz with a 95 \% confidence interval of [127, 162] Hz, suggesting an characteristic entangling time of 3.5 ms. This time scale is much longer than the system coherence and thus has negligible impact on the experiment. The on-off ratio is as high as $2.9\times10^4$. Another advantage of our scheme is that no extra control (e.g. flux bias\cite{zhang_tunable_2024} or cancellation microwave\cite{xiong_arbitrary_2022}) is required to suppress the static ZZ interaction. It is also inherently much lower than other fluxonium systems\cite{ficheux_fast_2021,dogan_two-fluxonium_2023,ding_high-fidelity_2023,lin_24_2024}. The low static ZZ rate benefits from the following aspects. Firstly, the computational states of fluxonium have small charge matrix elements, decoupling themselves from the high energy states. They behave more like spins, and therefore the static ZZ interaction of a fluxonium system is typically an order of magnitude smaller than that of transmon systems. Secondly, we make use of the multi-path coupling effect\cite{kandala_demonstration_2021,nguyen_blueprint_2022,ding_high-fidelity_2023,lin_24_2024}. The static ZZ interactions from the $J_\mathrm{c}$ terms and the $J_\mathrm{AB}$ term in Eq.~\ref{eq:Hamiltonian} cancel each other. Finally, the coupling strength of our system is weaker than that of other fluxonium experiments. With all the above, we achieve extremely low static ZZ interaction while maintaining the entangling ability of a 120-ns CZ gate.

\section{Bus time-domain measurements\label{sec:bus_time}}
\begin{figure}
    \centering
    \includegraphics[width=86mm]{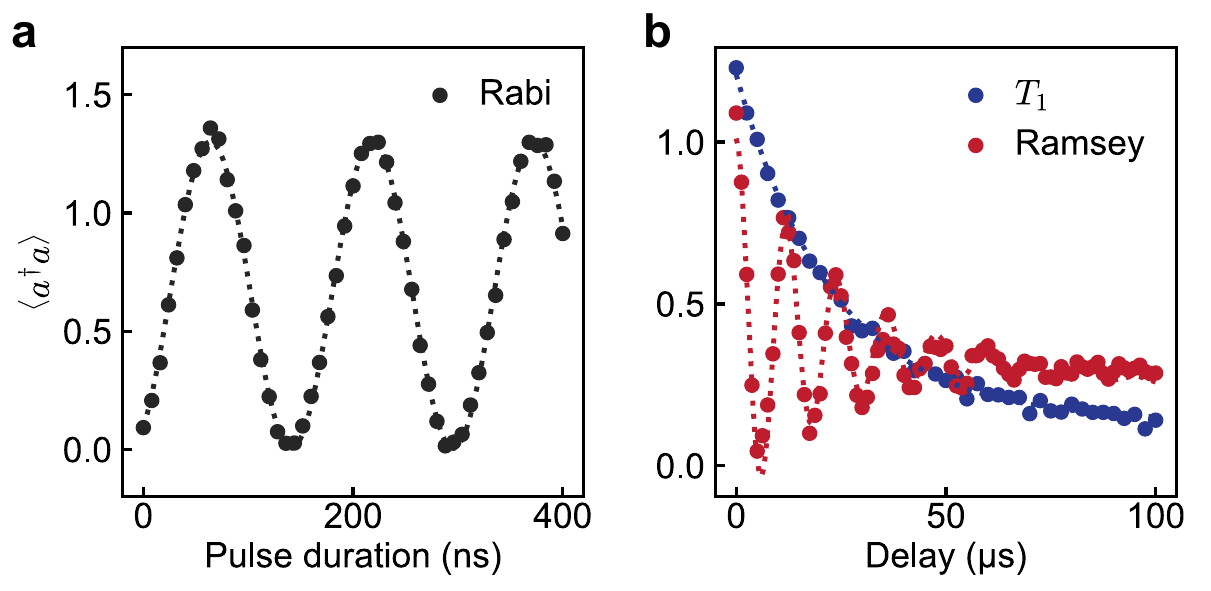}
    \caption{\textbf{Time-domain measurements of bus photons} (a) Rabi oscillation of the photon state. When the resonator is off-resonantly driven, the average photon number oscillates at the detuning frequency. (b) $T_1$ and $T_2$ of the photon state. We obtain $T_1=23.4\pm0.5~\mu\mathrm{s}$ and $T_2=22.2\pm1.1~\mu\mathrm{s}$ when there is almost one photon in the bus resonator.}
    \label{fig:bus_time}
\end{figure}

The photon calibration technique enables us to perform time-domain measurements on the photon coherent state. When off-resonantly driving the resonator with a square pulse with an amplitude of $\Omega$, the average photon number oscillates periodically at the detuning frequency $\Delta$ with a oscillation amplitude $|\Omega/\Delta|^2$ (Fig.~\ref{fig:bus_time}a). Similar to the qubit coherence measurement, we measure the photon energy relaxation time $T_1=23.4\pm0.5~\mu\mathrm{s}$ and decoherence time $T_2=22.2\pm1.1~\mu\mathrm{s}$. The data is presented in Fig.~\ref{fig:bus_time}b. The $T_1$ corresponds to a photon decay rate of $\kappa/2\pi=6.8~\mathrm{kHz}$ and quality factor $Q=\omega_r/\kappa=8.0\times10^5$. The photon lifetime is longer than that in a previous RIP gate experiment using 3D cavity\cite{paik_experimental_2016} owing to our high-quality tantalum film and surface treatment process\cite{place_new_2021,wang_towards_2022}.The photon decoherence time shows $T_2<2T_1$, which is probably due to the coupling between the bus and low-coherence fluxonium $\ket{3}$ states.

\section{Initialization and readout\label{sec:readout}}
\begin{figure}
    \centering
    \includegraphics[width=86mm]{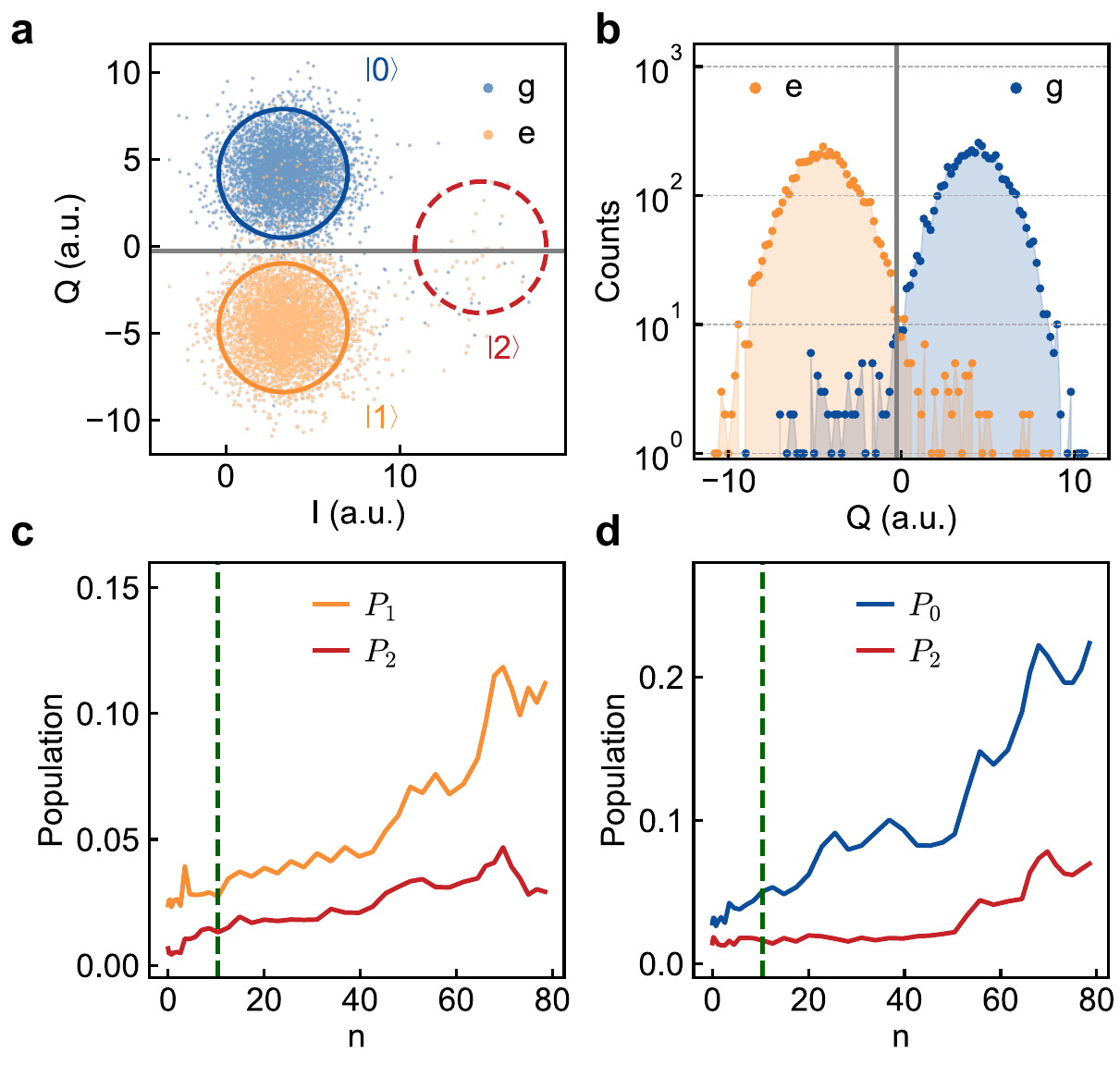}
    \caption{\textbf{Readout characterization of qubit A} (a) Single-shot measurement on qubit A when prepared at g state (blue dots) and e state (orange dots). The circles of state $\ket{0},\ket{1},\ket{2}$ represent the boundary when calculating the population using circular thresholds. The gray line in the middle represents the binary threshold. (b) Single-shot histogram on the Q quadrature. The vertical gray line indicates the binary threshold. (c) Measurement-induced state transition (MIST) versus readout photon number when qubit A is at g state. The photon number is calibrated through ac-Stark shift. The green dashed line indicates the photon number used in qubit readout ($n=10.4$). We observe measurement-induced transition to state $\ket{1}$ and $\ket{2}$. (d) MIST when qubit A is at e state. Population can also be transferred to state $\ket{0}$ and $\ket{2}$}
    \label{fig:readout}
\end{figure}

In this section, we discuss optimization and characterization of qubit reset and readout. Because fluxonium qubit frequency is lower than 1 GHz at half flux quanta, the thermal excitation is significant. We perform qubit initialization by simultaneously pumping fluxonium $\ket{1}-\ket{2}$ transition and $\ket{2,0}-\ket{0,1}$ transition (which is also known as the f0g1 transition) in the coupled system. During this process, the population is converted to readout resonator excitation $\ket{0,1}$ and then quickly decays back to $\ket{0,0}$. Here we call the initial state prepared with the above protocol the g state and prepare the e state by applying a $\pi$ pulse to the g state. Fig.~\ref{fig:readout} shows the demodulated signal of the single-shot measurement on qubit A. Apart from the two blobs corresponding to $\ket{0}$ state and $\ket{1}$ state, we also observe a small amount of data points fall into the blob of $\ket{2}$ state (Fig.~\ref{fig:readout}a). We obtain the blob centers and standard deviations by performing 2D Gaussian fit on the histograms. The qubit population can be extracted from the binary-thresholded outcomes. We can use the perpendicular bisector between the two blob centers as the demarcation threshold, as depicted with the gray line in Fig.~\ref{fig:readout}a,b. In this way, the $\ket{0}$ state population when the qubit is at g state is $P(0|g)=0.9826$, and the $\ket{1}$ state population for e state is $P(1|e)=0.9806$. The binary readout fidelity is $\mathcal{F}=[P(0|g)+P(1|e)]/2=0.9816$. We also execute two successive readout and obtain the readout repeatability or QND-ness $\mathcal{Q}=[P(0|0)+P(1|1)]/2=0.9728$. 

The binary-threshold method is simple and makes full use of all the data points, but it can introduce assignment error. For example, $\ket{0}$ state can be mistaken for $\ket{1}$ state (or vice versa) when the two blobs have significant overlap due to limited signal-to-noise ratio. Assignment errors also occur when there is leakage to higher states. In comparison, we then calculate the population using circular thresholds. As depicted with the blue and orange circles in Fig.~\ref{fig:readout}a, the radius is 2 times the blob standard deviation. By only counting the points inside the circles, we can curb the impact from blob overlap and eliminate the leakage to the high energy states. However, this is at the cost of lower data efficiency and more complicated discrimination process, which is potentially problematic in large-scale experiments. Discarding the 11 \% data points outside the circles, we obtain $P(0|g)=0.9900,P(1|e)=0.9883$, readout fidelity $\mathcal{F}=[P(0|g)+P(1|e)]/2=0.9891$, QND-ness $\mathcal{Q}=[P(0|0)+P(1|1)]/2=0.9840$. The circular-threshold method improves the readout performance significantly.

Alternatively, we can fit the $\ket{0}$, $\ket{1}$ and $\ket{2}$ state blob histogram with 2D Gaussian distribution, and calculate the population from the blob heights. This method avoids assignment error but does not apply to multi-qubit systems. The fitting gives $P(0|g)=0.9894,P(1|e)=0.9842$ and an average population of $P(2)=0.0042$ leaks to $\ket{2}$ state.

The above measurements demonstrate that fluxonium systems can achieve high readout fidelity and QND-ness after optimizing initialization and readout parameters. Its performance is close to state-of-the-art transmon device, proving the feasibility of replacing transmon with fluxonium as the basic unit in large-scale superconducting circuits.

A common potential problem in dynamics of the readout process is measurement-induced state transition (MIST), occuring in the case where a resonator is coupled to a multi-level system. Both theoretical and experimental explorations have been conducted in transmon as well as in fluxonium systems, revealing that it is one of the major limitations on superconducting qubit readout fidelity. In our system, we measure the fluxonium high energy state population to further explore MIST in fluxonium systems. We first send a pump pulse of the same length as the readout pulse (1700 $\mu\mathrm{s}$) to the readout resonator and measure the qubit population after the resonator photon is depleted. We scan the pump pulse amplitude and calibrate the steady state photon number in a separate ac-Stark shift measurement. The solid lines in Fig.~\ref{fig:readout}c,d represent the unwanted population of qubit A when it is prepared at g state and e state respectively. The population is extracted by performing the Gaussian fit. When the steady state photon is 0, the unwanted population is resulted from initialization infidelity and energy relaxation. For example, $\ket{2}$ state population in Fig.~\ref{fig:readout}d is higher than that in Fig.~\ref{fig:readout}c at zero photon, due to the thermal excitation process from $\ket{1}$ state to $\ket{2}$ state. In both cases, we observe the population undergoing MIST increases with photon number. Apart from the transition inside the computational space, $\ket{2}$ state population also increases significantly at high photon number. The readout pulse corresponds to a steady state of $n=10.4$, indicated by the green dashed line and the MIST effect is already noticeable in our readout process.

\section{Pulse optimization\label{sec:pulse_opt}}
\begin{figure}
    \centering
    \includegraphics[width=86mm]{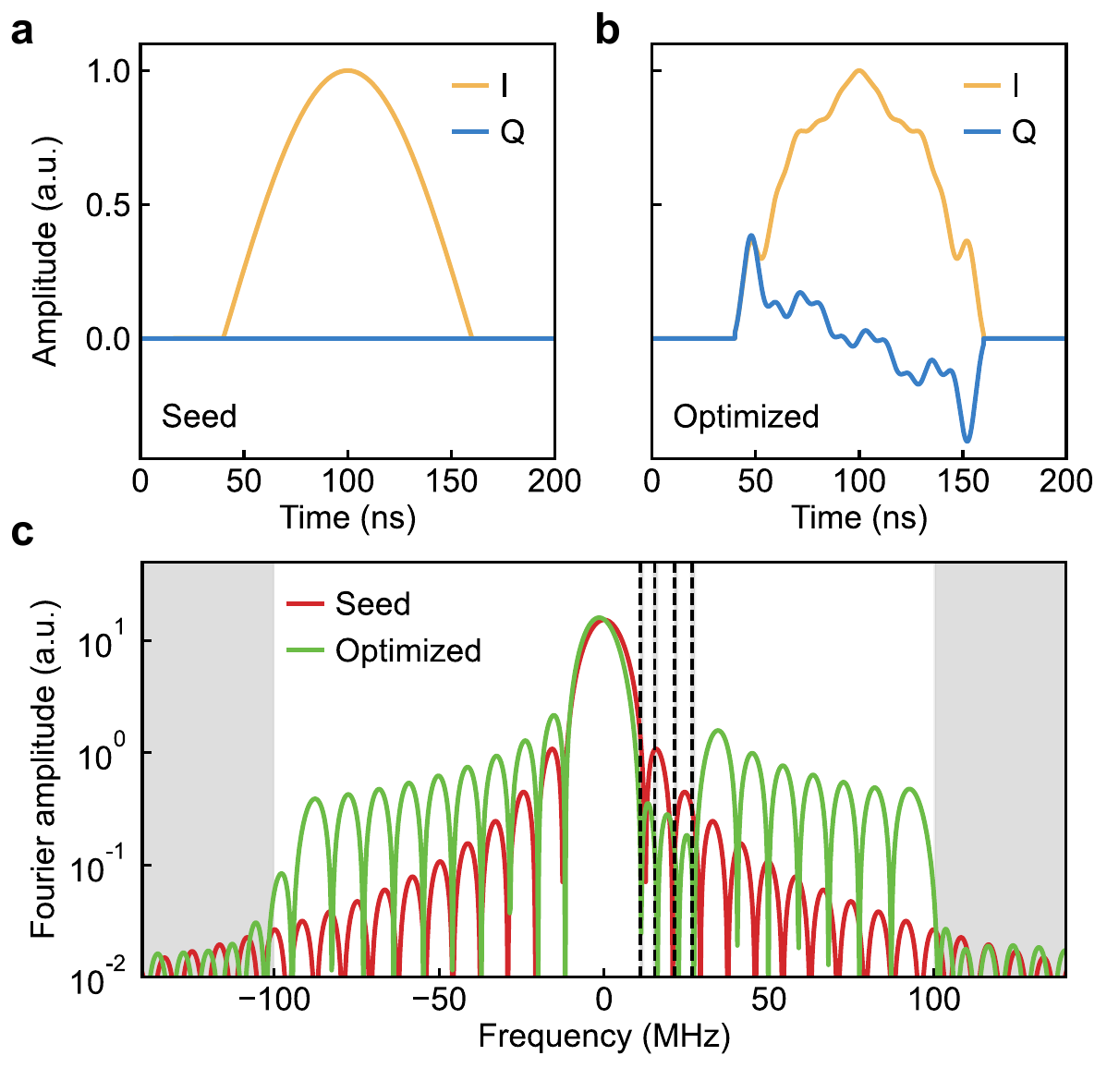}
    \caption{\textbf{Iterative spectrum engineering method} (a) I and Q quadrature of the seed pulse. We use a cosine envelop as the starting point of the optimization process. (b) I and Q quadrature of the optimized pulse shape. (c) Fourier spectrum before and after the optimization. The red curve is the Fourier amplitude of the seed pulse. The green curve is the Fourier amplitude after 100 iterations. The gray regime displays the Fourier-amplitude-suppression frequencies during iteration.}
    \label{fig:ise}
\end{figure}

In order to suppress residual photons, we use iterative spectrum engineering (ISE) method to optimize pulse shape. First, we need a seed pulse envelope $\Omega_0(t)$ as the starting point. Here the seed pulse is a 120-ns cosine pulse, as shown in Fig.~\ref{fig:ise}a. According to the target positions of the zero points $f_i$, we flip the sign of the pulse Fourier spectrum $F_0(f)$ within the region of $[f_i-0.5~\mathrm{MHz},f_i+0.5~\mathrm{MHz}]$ to suppress the corresponding Fourier amplitude in the following iteration. Likewise, we can also flip the sign in the region of $f<-100~\mathrm{MHz}$ and $f>100~\mathrm{MHz}$ to suppress the influence on transitions of large detuning. We then apply inverse Fourier transform on this new spectrum $F_1(f)$ and truncate the new pulse outside the 120-ns gate time. In this way, we complete one iteration and obtain a new pulse envelope $\Omega_1(t)$. The above process suppresses the unwanted Fourier components of the seed pulse $\Omega_0(t)$ while minimizing the deviation from the pristine spectrum and restricting the pulse to a fixed gate time. The method converges after 100 iterations in practice. The optimized pulse and its spectrum are plotted in Fig.~\ref{fig:ise}b,c. Our method can be combined with multi-derivative pulse shaping\cite{li_experimental_2024} to improve the efficiency in adjusting zero points while suppressing the overall Fourier amplitude at large detuning.

\section{Error budget estimation\label{sec:error}}
In this section, we introduce the numerical simulation for error budget estimation. For simplicity, we only include the fluxonium-bus coupling terms in the system Hamiltonian
\begin{equation}   H/h=\hat{H}_\mathrm{fA}+\hat{H}_\mathrm{fB}+\hat{H}_\mathrm{bus}+iJ_1\hat{n}_\mathrm{fA}(a^\dag-a)+iJ_2\hat{n}_\mathrm{fB}(a^\dag-a),
\label{eq:H_simu}
\end{equation}
where $J_1=98$ MHz and $J_2=94$ MHz in order to match the measured dispersive shifts.
\subsection{Unitary dynamics}\label{sec:simulations-unitary}

To diagonalize the Hamiltonian in Eq.~\ref{eq:H_simu}, we keep 8 lowest energy levels for fluxoniums and 30 energy levels for the bus. We then truncate the Hilbert space to a dimension of 800 when simulating the gate unitary dynamics. As the first step, we optimize the pulse shape to minimize the calculated residual photons following the method in Appendix~\ref{sec:pulse_opt}. The pulse parameters slightly deviate from those in the experiment, which can be attributed to the pulse distortion.  Notably, the optimized average residual photon number is $0.0059$, and is consistent with the experimental results. These residual photons can not be further eliminated in our simulation and we attribute them to the deviation from the ideal state $\ket{i}\otimes\ket{\alpha}$ due to the bus nonlinearity, where $\ket{i}$ stands for a qubits' state and $\ket{\alpha}$ represents a coherence state of the resonator. The measurement-induced state transitions of the gate process are negligible with the maximum leakage populations at the $10^{-8}$ level for all the four computational states. This ensures that the qubit population is within the computational space and the gate error can be corrected with error correction codes.

The state fidelity is calculated with $F(\hat{\sigma},\hat{\rho})=\mathrm{tr}\sqrt{\hat{\rho}^{\frac{1}{2}}\hat{\sigma}\hat{\rho}^{\frac{1}{2}}}$, where $\hat{\sigma}$ is the density matrix of the ideal final state and $\hat{\rho}$ is the simulated density matrix. The gate fidelity is estimated by averaging the state fidelities over 36 initial two-qubit states generated from the set of six initial single-qubit states $\{\ket{0},\ket{1},\frac{1}{\sqrt{2}}(\ket{0}\pm\ket{1}),\frac{1}{\sqrt{2}}(\ket{0}\pm i\ket{1})\}$. 
With the optimized pulse, we obtain a gate fidelity of 99.71$\%$ considering all the three bodies in the Hamiltonian, which is caused by the residual photons. If we only focus on the qubit Hilbert space by tracing out the bus degree of freedom, the gate fidelity reaches 99.87$\%$, implying that a part of the residual photons is not entangled with the qubit and therefore has less impact on the gate fidelity. 

\subsection{gate performance under decoherence}\label{sec:simulations-decoherence}

\begin{center}
\begin{table*}
\renewcommand{\arraystretch}{1.5}
\setlength{\tabcolsep}{5pt}
\begin{tabular}{  c |  c  c  c  c  c  c  c  c} \hline
    Qubit & $E_J$ (GHz) & $E_C$ (GHz) & $E_L$ (GHz) & $f_{01}$ (MHz) & $f_r$ (GHz) & $T_1~(\mu\mathrm{s})$ & $T_2^\mathrm{R}~(\mu\mathrm{s})$ &  $T_2^\mathrm{E}~(\mu\mathrm{s})$                      \\ \hline
    A  & 3.58 &1.03&0.54&349&6.964& 433 &72&99 \\ \hline
    B  & 3.91 &1.01&0.52&267&6.997& 113 &17&39 \\ \hline
\end{tabular}
\caption{\label{tab:coherence} Fluxonium parameters. In this table, we show the circuit parameters ($E_J$, $E_C$ and $E_L$), 0-1 transition at half flux quanta $f_{01}$, readout frequency $f_r$, energy relaxation time and coherent times of the two fluxoniums. The Josephson energy $E_J$, charging energy $E_C$ and inductive energy $E_L$ are extracted from spectrum fit. The coherent times measured with Ramsey ($T_2^\mathrm{R}$) and spin-echo ($T_2^\mathrm{E}$) experiments are much shorter than $2T_1$. It is likely limited by the dephasing from the flux noise, as qubit B has worse coherence, narrower flux sweet spot and lower qubit frequency.}
\end{table*}
\end{center}

The gate fidelity from the unitary dynamics simulation is higher than the measured value, indicating decoherence process is the dominating error source. The qubit coherence times are listed in Table~\ref{tab:coherence}. The corresponding incoherent error can be described with\cite{ding_high-fidelity_2023, li_realization_2024}
\begin{equation}
    \epsilon_Q=\frac{2t_g}{5}\left(\frac{1}{T_1^A}+\frac{1}{T_\phi^A}+\frac{1}{T_1^B}+\frac{1}{T_\phi^B}\right).
\end{equation}
For a 120-ns CZ gate, we obtain $\epsilon_Q=0.21 \%$. Another source of incoherent errors is the bus photon. To estimate the bus-induced decoherence, we consider a Lindblad master equation
\begin{equation}
    \dot{\hat{\rho}}=-\frac{i}{h}[\hat{H},\rho]+\sum_k[\hat{L}_k\hat{\rho}\hat{L}_k^\dag-\frac{1}{2}(\hat{L}_k^\dag\hat{L}_k\hat{\rho}+\hat{\rho}\hat{L}_k^\dag\hat{L}_k)].
\end{equation}
The relaxation and dephasing of the bus photon can be described with the collapse operators:
\begin{equation}
\begin{split}
        \hat{L}_1=\sqrt{\Gamma_1}a,\\
        \hat{L}_2=\sqrt{2\Gamma_\phi} a^\dag a,
\end{split}
\end{equation}
where $\Gamma_1=1/T_1 $ and $\Gamma_\phi=1/T_2-1/2T_1$ are extracted from the experiments described in Appendix~\ref{sec:bus_time}. We first ignore the $\hat{L}_2$ term and use a general closed-form expression to calculate relaxation-induced dephasing. The off-diagonal element $|C_{ij}(t)|$ of the two-qubit density matrix can be described as 
\begin{equation}
    {\rm Re}\left\{C_{ij}(0){\rm exp}\left[\Gamma_1\int_0^t d\tau (\alpha_i\alpha_j^*-\frac{1}{2}|\alpha_i|^2-\frac{1}{2}|\alpha_j|^2)\right]\right\},
\end{equation}
where $i, j$ label the computational states, and the photon trajectories $\alpha_{i,j}(t)$ are extracted from the unitary evolution. We multiply these dephasing factors to the corresponding density matrix elements of the unitary dynamics to estimate the decoherence error in the presence of the photon loss. As a result, the relaxation-induced dephasing error on the two-qubit unitary is 0.03~\%, indicating that the bus $T_1$ has negligible effect on the qubit coherence. Next, we use the Monte Carlo Solver provided by Qutip \cite{QuTiP} to simulate the effect from both the photon relaxation and the dephasing with a reduced Hilbert space that has a dimension of 280 (truncated from the full Hilbert space with 5 levels of a single fluxonium and 13 levels of the bus). As a result, 
the two-qubit unitary fidelity is $99.67\%$ after tracing out the bus degree of freedom. Therefore, the total gate error including single-qubit decoherence is 0.54~\% which is close to the measured error of 0.63~\%. The discrepancy can be explained with coherence fluctuation and fitting errors. In summary, our simulation shows that the photon dephasing plays a significant role in the error budget. The gate fidelity can be readily improved by means of increasing the coupling and reducing the gate time.

\end{document}